\documentclass[aps,prd,floats,nofootinbib,preprintnumbers,twocolumn]{revtex4}

\usepackage[utf8]{inputenc}
\usepackage{mathrsfs}
\usepackage{amssymb}
\usepackage{amsmath}
\usepackage{feynmp}
\usepackage{slashed}
\usepackage{latexsym}
\usepackage{graphicx}
\usepackage{verbatim}
\usepackage[normalem]{ulem}

\newcommand{\be}{\begin{equation}}
\newcommand{\ee}{\end{equation}}
\newcommand{\ba}{\begin{array}}
\newcommand{\ea}{\end{array}}
\newcommand{\bea}{\begin{eqnarray}}
\newcommand{\eea}{\end{eqnarray}}
\newcommand{\balg}{\begin{align}}
\newcommand{\ealg}{\end{align}}
\newcommand{\bit}{\begin{itemize}}
\newcommand{\eit}{\end{itemize}}
\newcommand{\trm}[1]{\textrm{#1}}

\newcommand{\mcl}[1]{\mathcal{#1}}
\newcommand{\mbb}[1]{\mathbb{#1}}
\newcommand{\msc}[1]{\mathscr{#1}}

\newcommand{\gfl}{\mcl{G}_{fl}}

\newcommand{\Mpc}{\trm{\Mpc}}
\newcommand{\yr}{\trm{\yr}}
\newcommand{\eV}{\trm{\eV}}

\newcommand{\tr}[1]{\trm{Tr}\left[ {#1} \right]}

\begin{document}

\title{Flavorspin}

\author{Jeffrey M. Berryman and Daniel Hern\'{a}ndez}
\affiliation{Northwestern University, Department of Physics \& Astronomy, 2145 Sheridan Road, Evanston, IL~60208, USA}

\begin{abstract}
We propose that the flavor structure of the quark sector of the Standard Model is determined by a vectorial $SU(2)$ flavor symmetry, which we dub Flavorspin, under which quarks transform as triplets. The fundamental Yukawa couplings are real and $CP$ violation is not directly linked to the breaking of the flavor symmetry. A $CP$-conserving scenario is naturally defined with the feature that the Yukawa spurions are completely determined in terms of the masses and mixings. $CP$ violation may be introduced with negligible impact on low-energy observables other than generating a large $\delta_{CP}$ in the Standard Model mixing matrix. The scale of flavor-symmetry violation must be large in order to prevent sizable Flavor Changing Neutral Currents, which can be partially suppressed if Flavorspin is a residual symmetry of a larger flavor group.

\end{abstract}

\maketitle

\setcounter{equation}{0}
\setcounter{footnote}{0}


\section{Introduction}


An understanding of the Standard Model (SM) flavor structure has remained elusive for decades. The masses of the three families of fermions, with the possible exception of the neutrinos, show a distinctly hierarchical pattern. Mixings between the two weak isodoublet components are small in the quark sector but are large in the lepton sector. In this letter we argue that the flavor structure in the quark sector of the SM can be well accommodated within a global $SU(2)$ flavor symmetry under which all quarks transform as triplets. As we will see, some features of the SM flavor structure can be \emph{understood} if such a symmetry is in place.

The terms of relevance in the SM Lagrangian are the quark Yukawa terms which are given by
\be
\msc{L}_{Yuk}  = - \overline{Q}_L Y_U U_R \widetilde{H} - \overline{Q}_L Y_D D_RH + \trm{h.c.} \label{lag}
\ee
where $Q_L$ stands for the left-handed (LH) quark doublets and $U_R,\, D_R$ denote the up and down right-handed (RH) singlets, respectively. $H$ is the Higgs doublet with $\tilde{H} = i\tau_2H^*$ and $\langle H\rangle = v/\sqrt{2}$ and $v = 246$ GeV.

When the quark Yukawa couplings are set to zero, the Lagrangian in Eq.~\eqref{lag} gains a very large global flavor symmetry,
\be
\mcl{G}_0 = U(3)^3 \equiv SU(3)_Q \otimes SU(3)_D \otimes SU(3)_U \otimes U(1)^3 \,, \label{flavor-group}
\ee
where the $U(1)$ factors correspond to baryon number, hypercharge and Peccei-Quinn symmetry, and $SU(3)_{Q,D,U}$ stand for independent unitary rotations that transform the $Q_L$-, $U_R$- and $D_R$-type quarks amongst themselves. The simple $SU(3)$ factors and the Peccei-Quinn $U(1)$ are broken once the Yukawas are turned on. 

Within the spurion interpretation, charges under a flavor group $\mcl{G}_{fl}$ are formally assigned to the quark fields and Yukawa couplings so as to render the Lagrangian in Eq.~\eqref{lag} invariant. One may postulate that if the SM is regarded as an effective theory, then $\gfl$ should also be respected by any nonrenormalizable operators made of SM fields.

A crucial point in this work is that the equivalence $\mcl{G}_{fl} \equiv \mcl{G}_0$ is \emph{not} necessary. Formal invariance under $\mcl{G}_0$ can emerge \emph{accidentally} as a consequence of $\mcl{G}_{fl}$-invariance. Any group $\mcl{G}_{fl} \subset \mcl{G}_0$ that has irreducible 3-dimensional (3D) representations can do the trick. In particular, a vectorial $SU(2)_V$ symmetry is minimal. Such a global horizontal symmetry was first discussed in \cite{wilczekzee} -- the local version had been discussed previously in \cite{yanagida} -- and subsequently in many papers\cite{others, others1, others2}. Here, we propose a new approach to explore formal invariance under $\gfl = SU(2)_V$. We dub this scenario Flavorspin.


\section{Flavorspin}


We assume that identically-charged quarks belong to triplet representations of $SU(2)_V$. In order for the SM to be formally $\gfl$-invariant, Yukawa spurions must transform as a 5D real symmetric traceless tensor or as a 3D real antisymmetric tensor. Terms proportional to the identity are also allowed by the symmetry. Thus, the most general $Y_U$ and $Y_D$ have the form
\begin{align}
Y_U & = e^{i\eta_5}Y_5 + e^{i\eta_3} Y_3  + \frac{y}{3} \mbb{I} \,, \label{yu} \\
Y_D & = \alpha \left( e^{i\zeta_5} Y_5  + \beta_3 e^{i\zeta_3} Y_3 +  \beta_d \frac{y}{3} \mbb{I} \right) \,, \label{yd}
\end{align}
where $\alpha,\,\beta_3$ and $\beta_d$ are real numbers and $Y_5$ and $Y_3$ are \emph{real} flavor spurions  
\be
Y_5\sim 5_{fl} \,,\quad Y_3 \sim 3_{fl} \,.
\ee
with $Y_5 = Y_5^T$, $Y_3 = -Y_3^T$. Under $\gfl$, quarks transform as
\be
Q_L \rightarrow \mcl{O}Q_L\,,\quad U_R \rightarrow \mcl{O}U_R\,,\quad D_R \rightarrow \mcl{O}D_R\,,
\ee
where $\mcl{O}$ is a $3\times 3$ orthogonal matrix. The flavor spurions transform as 
\be
Y_{5,3} \rightarrow \mcl{O}Y_{5,3} \mcl{O}^T \,. \label{sp-transf} 
\ee
Notice that in Eqs.~(\ref{yu}, \ref{yd}) we have allowed for complex coefficients. Moreover, we have absorbed the absolute value of the coefficients in $Y_U$ into the definitions of $Y_5$ and $Y_3$. We have used the phase redefinitions at our disposal to render the singlet terms real.

The flavor spurions are strictly real, so the complex coefficients must be responsible for any violation of $CP$ symmetry. $CP$ violation appears to be due to physics unrelated to the violation of $\gfl$. We start by analyzing the case in which $CP$ is conserved, i.e., the case in which all coefficients in Eqs.~(\ref{yu}, \ref{yd}) are real. As we will argue below, $CP$-conserving Flavorspin may be a good approximation to the more realistic $CP$-violating case.

\subsection{$CP$-Conserving Scenario}

With the phases $\eta_{5,3}$ and $\zeta_{5,3}$ set to zero the Yukawas
\be
Y_u = Y_5 + Y_3 + \frac{y}{3}\mbb{I}\,,\quad Y_d =  \alpha \left(Y_5 + \beta_3 Y_3 + \beta_d \frac{y}{3}\mbb{I}\right)\,, \label{yuyd2}
\ee
are both real with $\tr{Y_u} = y$, $\tr{Y_d} = \beta_dy$. The Yukawas $Y_u$ and $Y_d$ can be diagonalized by biorthogonal transformations
\be
\mcl{O}_uY_u\tilde{\mcl{O}}^T_u = M_u \,,\quad \mcl{O}_dY_d\tilde{\mcl{O}}^T_d = M_d \,, \label{basis-change}
\ee
where 
\be
M_u = \trm{diag}\{m_u,\, m_c,\,m_t \} \,,\quad  M_d = \trm{diag}\{m_d,\, m_s,\,m_b \} \,.
\ee

The mixing matrix $\mcl{O}_{CKM}$ is given by
\be
 \mcl{O}_{CKM}  = \mcl{O}_u\mcl{O}_d^T
\ee
and is orthogonal in the $CP$-conserving scenario. Notice that a change of basis such as the one in Eq.~\eqref{basis-change} takes the Yukawas to the physical basis but it does not preserve the symmetry properties of $Y_5$ and $Y_3$, since it is not a member of the flavor group $\gfl$.

On the other hand, changes of bases by elements of $\gfl$ do not alter the symmetry properties of $Y_5$ and $Y_3$. We can use the group transformations in Eq.~\eqref{sp-transf} to go to a basis  $\overline{Y}_5$, $\overline{Y}_3$ such that $\overline{Y}_5$ is diagonal and traceless:
\be
\bar{Y}_5 = \left( \begin{array}{ccc}
a & 0 & 0 \\ 0 & b & 0 \\ 0 & 0 & -a-b 
\end{array} \right) \,,\quad 
\bar{Y}_3 = \left( \begin{array}{ccc}
0 & X & Y \\ -X  & 0 & Z \\ -Y & -Z & 0
\end{array} \right) \,.
\ee
We call this the \emph{symmetric} basis.

It is illustrative to do a naive parameter counting. Let us collectively denote the parameters in the Yukawa coefficients in Eq.~(\ref{yuyd2}) by $\vec{\,p}$. The symmetric basis makes it clear that there are 9 independent parameters in $\vec{\,p}$:
\be
\vec{\,p} = \{a,\,b,\,X,\,Y,\,Z,\,\alpha,\,\beta_3,\,\beta_d,\,y\}\,.
\ee
This exactly matches the number of physical observables in the quark sector of the SM model with no $CP$ violation -- 6 masses and 3 mixing angles -- that we collectively denote by $\vec{\,o}$. Previous attempts to address the Flavor Puzzle by means of a vectorial $SU(2)$ focused on devising ansatze for $Y_5$ and $Y_3$ that yielded the desired values for the masses and mixings \cite{wilczekzee}. However, in $CP$-conserving Flavorspin we naively expect to be able to extract the Yukawas and the coefficients in Eqs.~(\ref{yu}, \ref{yd}) directly from the low-energy observables.

\begin{figure}[t!]
\includegraphics[width=85mm]{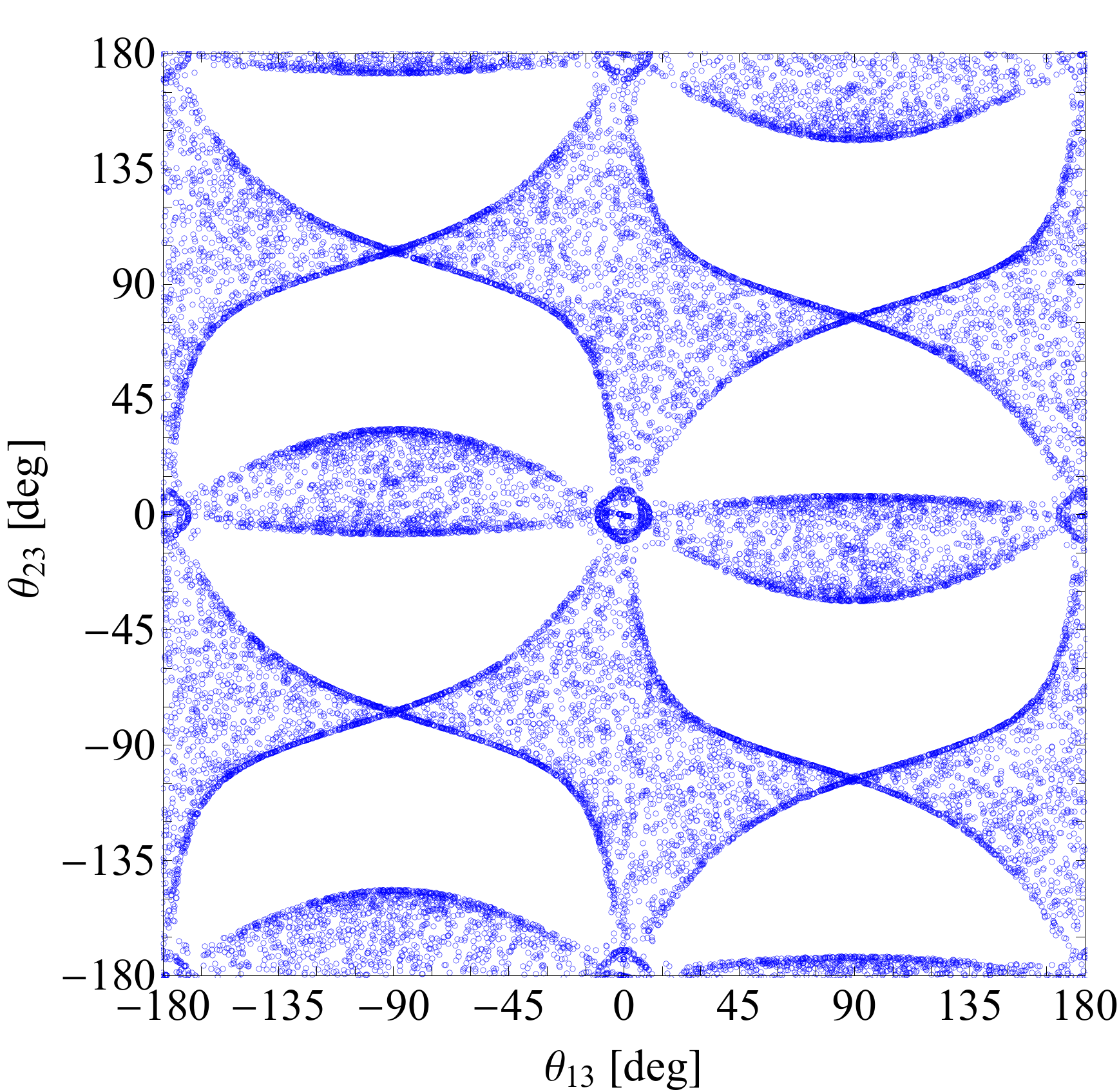}
\caption{The division of the $\theta_{13}\theta_{23}$-plane into regions where solutions exist (dotted) and regions of no solution (blank) for all other physical parameters fixed at the $\vec{\,o}_{SM}$ values. \label{th13th23}}
\end{figure}
The equations linking the physical observables $\vec{\,o}$ with the parameters $\vec{\,p}$ can be obtained as follows. First, use $\gfl$ transformations to rotate to yet another basis $Y_u'$ and $Y_d'$ such that
\begin{align}
Y_u' & = \mcl{O}_uY_u\mcl{O}_u^T = M_u\mcl{O}'_u \,, \\
Y_d' & = \mcl{O}_uY_d\mcl{O}_u^T = \mcl{O}_{CKM}M_d\mcl{O}'_d  \,.
\end{align}
Defining $Y_{5,3}' \equiv \mcl{O}_uY_{5,3}\mcl{O}_u^T$ and using Eq.~\eqref{yuyd2} we obtain
\begin{align}
Y_5' & = \frac{\alpha \beta_3 (Y_u'-\tr{Y_u'}) - (Y_d'-\tr{Y_d'})}{\alpha(\beta_3 - 1)} \, ,\label{y5'} \\
Y_3' & = \frac{\alpha (Y_u' -\tr{Y_u'}) - (Y_d'-\tr{Y_d'})}{\alpha(1 - \beta_3)} \,. \label{y3'}
\end{align}
By demanding that $Y_5'$ and $Y_3'$ satisfy the symmetry conditions
\be
Y_5' - Y_5'^T = 0\,,\quad Y'_3 + Y_3'^T = 0\,, \label{syseqs}
\ee
we obtain a set of 8 equations -- 5 equations from $Y_3'$ and 3 from $Y_5'$ -- with 8 unknowns - $\alpha$, $\beta_3$ and the 6 angles in $\mcl{O}'_u$ and $\mcl{O}'_d$. The problem is reduced to the technical problem of solving a system of nonlinear equations. The inputs for this system are the masses and mixing angles in $M_u$, $M_d$ and $\mcl{O}_{CKM}$. Solving the system yields $Y_u'$ and $Y_d'$ from which it is simple to find $\vec{\,p}$ by transforming to the symmetric basis.

We give an overview of the results we obtain, a detailed account of which will be given elsewhere \cite{bigpaper}. It is convenient to use a reference point in the space of observables which we call $\vec{\,o}_{SM}$, corresponding to the SM best-fit values for the masses and mixings. We use the PDG values \cite{PDG} for the masses and the mixings in the standard parametrization with no $CP$ violation \cite{standardparam}. In this approximation, we ignore that the quark masses are determined at different energy scales.

\begin{figure}[t!]
\includegraphics[width=80mm]{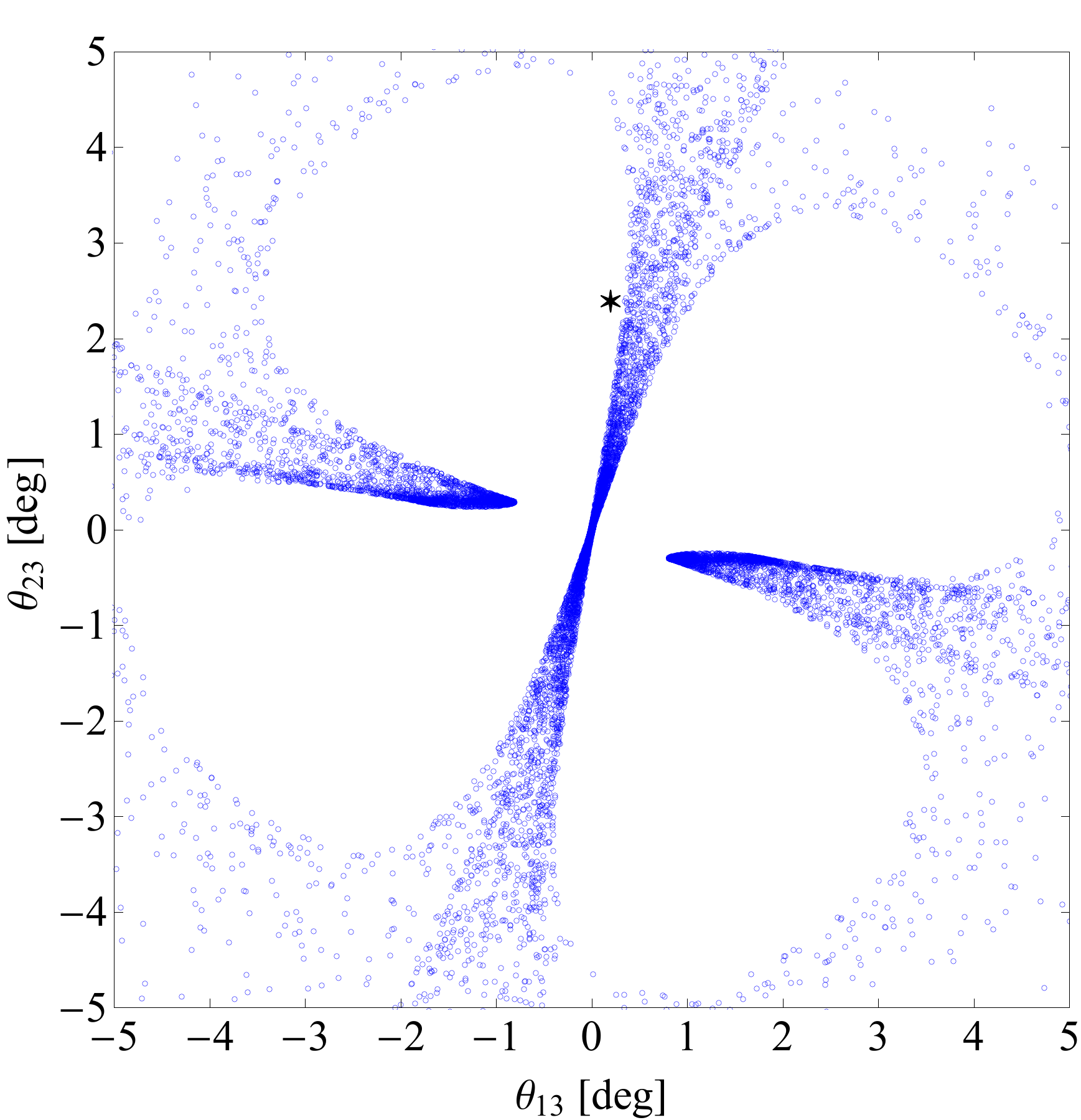}
\caption{A zoomed in region of Fig.~\ref{th13th23} but with $\theta_{13},\theta_{23} \in [-5^\circ,5^\circ]$. The black star corresponds to $\vec{\,o}_{SM}$. \label{th13th23-zoom}}
\end{figure}

\begin{center}
\begin{table}[h]
\begin{tabular}{c | c | c}
$\vec{\,o}$ & $\vec{\,o}_{SM}(\mu)$  &  $\vec{\,o}'_{SM}$ \\
\hline
\hspace{.2cm} $m_t$ \hspace{.2cm} & 160.0 GeV (160.0 TeV)  &  \hspace{.2cm} idem   \hspace{.2cm} \\
$m_c$ & 1.275 GeV (1.275 GeV) & idem \\
$m_u$ & \hspace{.2cm} 2.3 $\times 10^{-3}$ GeV (2 GeV) & idem \\
$m_b$ & 4.18 GeV (4.18 GeV) & idem\\
$m_s$ & 0.095 GeV (2 GeV)& idem\\
$m_d$ & 4.8 $\times 10^{-3}$ GeV (2 GeV)& idem \\
$\theta_{12}$ & 13.04º & idem \\
$\theta_{23}$ & 2.38º & idem \\
$\theta_{13}$ & 0.201º & 0.5º \\
\end{tabular}
\caption{Values of the observable parameters for the reference points $\vec{\,o}_{SM}$ and $\vec{\,o}'_{SM}$. Quark masses are estimated at the scale $\mu$.} 
\label{table1}
\end{table}
\end{center}

Due to the nonlinear character of Eq.~\eqref{syseqs}, we expect to find large swaths of parameter space $\vec{\,o}$ for which no $\vec{\,p}(\vec{\,o})$ solution exists.  We exemplify this in Fig.~\ref{th13th23}. The dotted regions correspond to the patches in the $\theta_{13}\theta_{23}$-plane where solutions are found when the rest of the observables are kept fixed at the $\vec{\,o}_{SM}$ values\footnote{The fundamental cell in the $\theta_{13}\theta_{23}$-plane in Fig.~\ref{th13th23} is given by the ranges $\theta_{13}\in [-\pi/2,\pi/2]$ and $\theta_{23} \in [0,\pi/2]$ as it should be for the SM with no $CP$ violation.}. There are large regions in which no solutions exist. A necessary condition for the $CP$-conserving Yukawa structure imposed by Flavorspin to be a good approximation to the SM is for $\vec{\,o}_{SM}$ not to lie deep within the regions of no solutions. 

\begin{center}
\begin{table}[t]
\begin{tabular}{c | c}
$\vec{\,p}$ &  $ \vec{\,p}'_{SM}$ \\
\hline
\hspace{.2cm} $a$ \hspace{.2cm} & \hspace{.2cm} $0.26245$ \hspace{.2cm} \\
$b$ & $0.33141$  \\
$X$ & \hspace{.2cm} $ -1.9152 \times 10^{-3} $ \hspace{.2cm} \\
$Y$ & 1.0436 $\times 10^{-2}$ \\
$Z$ & $-0.23366$ \\
$\alpha$ & $2.5401 \times 10^{-2}$ \\
$\beta_3$ & 1.2041 \\
$\beta_d$ & 1.0002\\
$y$ & -0.26160 \\
\end{tabular}
\caption{Example of solution for Eq.~\eqref{syseqs} for $\vec{\,o} = \vec{\,o}_{SM}'$}
\label{tab2}
\end{table}
\end{center}

Fig.~\ref{th13th23-zoom} shows a zoom of Fig.~\ref{th13th23} in the SM region of interest for these parameters. Fine detail is apparent, in particular the appearance of a four-leaf-clover-like region of no solutions around vanishing $\theta_{13}$ and $\theta_{23}$. The black star in the figure corresponds to $\vec{\,o}_{SM}$. The figure makes it clear that although there is no solution for $\vec{\,o}_{SM}$, it lies remarkably close to the border between the regions where solutions are found and those where there are none.

In order to study solutions in the vicinity of $\vec{\,o}_{SM}$,  we define the set of observables $\vec{\,o}'_{SM}$  for which solutions exist. The values of the observables are chosen in $\vec{\,o}'_{SM}$ to coincide with those in $\vec{\,o}_{SM}$, except for a small shift in $\theta_{13}$. This is detailed in Tab.~\ref{table1}. For any point $\vec{\,o}$ in the region of solutions, we expect degeneracies in the solution space;  indeed, we find there are several solutions $\vec{\,p}_i(\vec{\,o}_{SM}')$ for Eq.~\eqref{syseqs}. Some of these degeneracies are consequences of symmetries of the equations themselves \cite{bigpaper}. Barring these trivial degeneracies and focusing on the $\alpha,\beta_3>0$ quadrant, we find an 8-fold degeneracy in the parameter space $\vec{\,p}$. An example of solution is given by $\vec{\,p}'_{SM} \in \{\vec{\,p}_i(\vec{\,o}_{SM}') \}$ detailed in Tab. 2.

Note that the choice of $\vec{\,o}'_{SM}$, as well as the choice of $\vec{\,p}'_{SM}$ from the finite set of solutions to Eq.~\eqref{syseqs}, are arbitrary and purely for means of illustration. Moreover, the computation of the $\vec{\,p}'_{SM}$ coefficients from  $\vec{\,o}'_{SM}$ should be considered only as an approximation. In particular, the analysis thus far does not take $CP$ violation into account, nor does it take into account renormalization effects that presumably alter the connection between $Y_u$ and $Y_d$ expressed in Eqs.~(\ref{yu}, \ref{yd}). 

We have kept up to 5 digits of precision in Tab.~\ref{tab2} which we find is the minimum necessary to solve accurately for $\vec{\,o}_{SM}'$. This makes sense since $m_u/m_t \sim 10^{-5}$, and we find that indeed the up quark mass is the most sensitive parameter in  $\vec{\,o}_{SM}'$ to changes in  $\vec{\,p}'_{SM}$. A complete analysis on the stability of the solutions will be presented in \cite{bigpaper}.


For the solution $\vec{\,p}'_{SM}$, most of the couplings in $Y_5$ and $Y_3$ are natural in the sense that they are $\mathcal{O}(1)$. The hierarchies that do exist are milder than those of the SM. In particular, the singlet term $y$ is $\mathcal{O}(1)$ in contrast to \cite{wilczekzee} where it was assumed that $y$ vanishes\footnote{This assumption leads to a light top quark which was acceptable at the time.}. 
Another important feature is that $\beta_3, \,\beta_d \sim 1$. This is, in fact, common to all solutions, and it is a consequence of the fact that $\mcl{U}_{CKM} \sim \mbb{I}$; no mixing implies exact equality $\beta_3 = \beta_d = 1$. Adopting a top-down perspective, it suggests that the small mixing in the quark sector may be due to $\gfl$ being a subgroup of a larger symmetry group that enforces $\beta_3 = \beta_d = 1$ at some higher energy scale. Such relation would occur if, for instance, the LH and RH quarks were initially charged under a horizontal $SU(2)_L\otimes SU(2)_R$ flavor group with the presence of only one Yukawa spurion transforming according to the $(3,\,3)$ representation. In that case, at some scale $\Lambda_0$, $Y_u = \alpha Y_d$ would hold. Below $\Lambda_0$, the symmetry would break according to the patter $SU(2)_L\otimes SU(2)_R \rightarrow SU(2)_V$ would take place and the relation $Y_u = \alpha Y_d$ would receive corrections.

Since the fundamental Yukawas are obtained purely from the LH mixing, a hypothetical RH mixing matrix is completely determined in $CP$-conserving Flavorspin. Mixing among RH fermions is not observable in the SM but it is in Left-Right symmetric extensions.  This scenario could be ruled out if Left-Right symmetry is confirmed and the observed RH mixing matrix does not coincide with the Flavorspin prediction. 

\subsection{$CP$-Violating Scenario}

How does the scenario change when $CP$ violation is accounted for? $CP$ symmetry is generically not conserved if the phases in Eqs.~(\ref{yu}, \ref{yd}) are nonzero. Since $CP$ violation manifests in the SM through only  one observable, it is evident that we cannot solve uniquely for the parameters $\vec{\,p}$  as before. Nonetheless, it is possible to assess the impact on the physical parameters of adding the phases $\eta_{5,3}$ and $\zeta_{5,3}$ individually. Starting from a solution such as $\vec{\,p}'_{SM}$, we keep $Y_5$ and $Y_3$ fixed, and turn on one of the phases $\eta_{5,3}$ and $\zeta_{5,3}$. It is possible to work backwards and find how the physical observables vary when each of these phases runs through the interval $[-\pi,\,\pi]$.

We profit from the following observation: the masses of the light quarks are very sensitive to the phases $\eta_5$, $\eta_3$ and $\zeta_5$ for all solutions $\vec{\,p}_i(\vec{\,o}_{SM}')$ \cite{bigpaper}. Given that $CP$ violation comes from physics that is unrelated to the breaking of flavor symmetry, if any of these 3 phases is different from zero, either that phase has to be negligibly small or a fine-tuned cancellation must occur in order to ensure that the up and down quarks are light. If we demand that the values of $m_u$ and $m_d$ should be stable with respect to the introduction of phases, the only possible way to introduce $CP$ violation is through $\zeta_3$. 

\begin{figure}[t!]
\includegraphics[width=85mm]{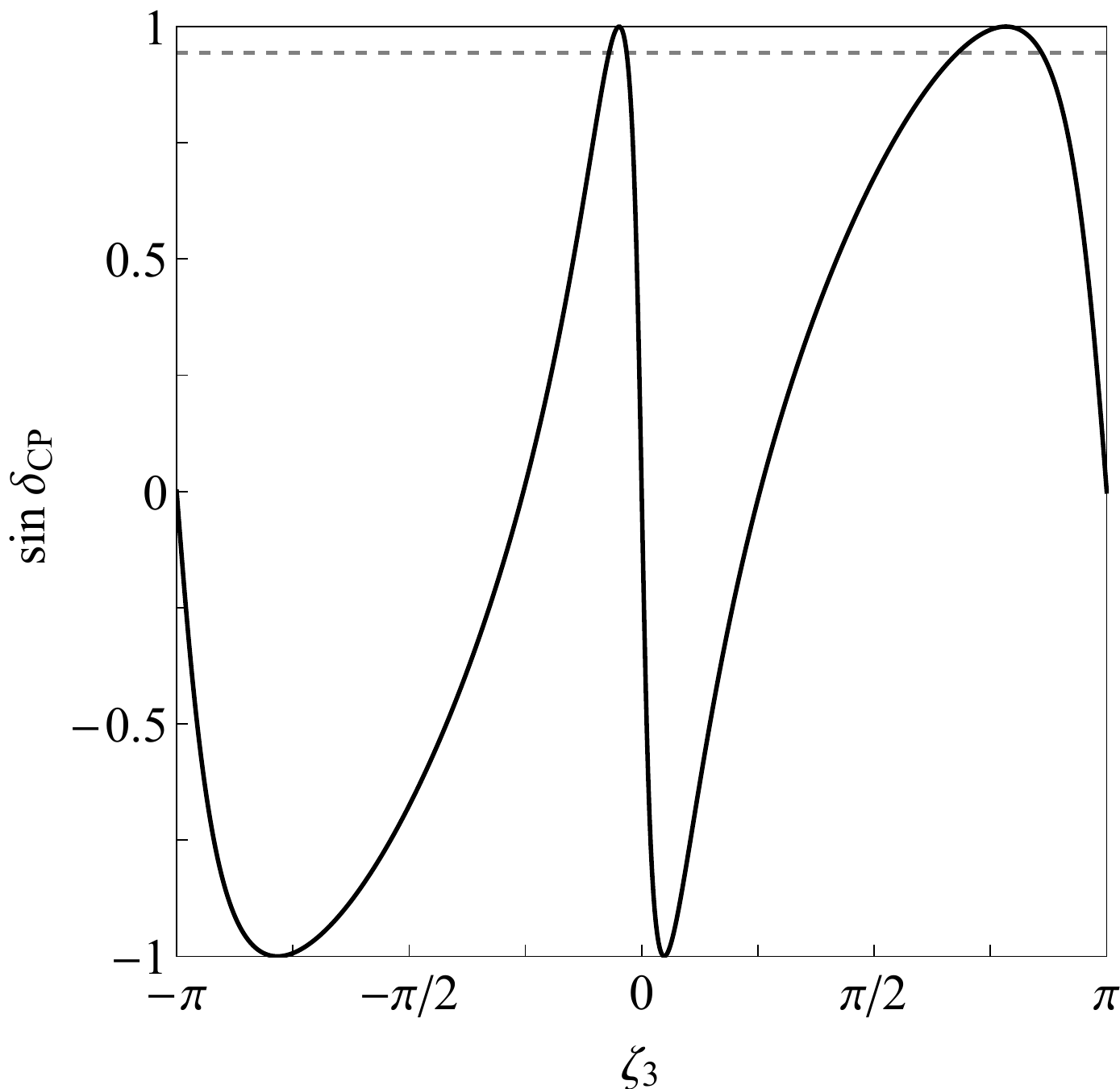}
\caption{The sine of the $CP$-violating phase $\delta_{CP}$ as a function of $\zeta_3$ for $\vec{\,p}_1(\vec{\,o}'_{SM}) + \zeta_3$. The horizontal dashed line corresponds to the value of $\delta_{CP}$\label{deltacp-zeta} reported by the PDG \cite{PDG}, $\sin \delta_{CP} = 0.944$.}
\end{figure}

Fig.~\ref{deltacp-zeta} shows the SM $CP$-violating phase $\delta_{CP}$ obtained for varying values of $\zeta_3$ using the solution $\vec{\,p}'_{SM}$. Large values of $\delta_{CP}$ can be achieved for relatively small values of $\zeta_3$. We find that no other SM parameter is appreciably perturbed by the small values of $\zeta_3$ required to reproduce $\delta_{CP}$ \cite{bigpaper}. We conclude that there are $\vec{\,p}$-solutions that, to a good approximation, reproduce all the SM observables. Moreover, it is possible to do so with a relatively small fundamental source of $CP$ violation. This justifies our previous claim that the $CP$-conserving scenario can be a good approximation 

We remark, however, that this is not a universal feature. Other solutions exist for which the impact of introducing $\zeta_3$ on $\delta_{CP}$ is not so dramatic. Nonetheless, for the specific case of $\vec{\,p}'_{SM}$, $CP$-conserving Flavorspin constitutes an excellent approximation to the SM.

\subsection{Flavorspin and Minimal Flavor Violation}

Although we believe Flavorspin is interesting in its own right, it is worthwhile to establish whether this scenario allows for the scale at which new physics may appear to be lowered -- that is, to establish whether it constitutes a viable alternative to Minimal Flavor Violation (MFV) \cite{georgichivukula}. In Flavorspin, $Y_5$ and $Y_3$ play the role of fundamental spurions, as opposed to $Y_U$ and $Y_D$ in the MFV framework \cite{MFVmain}.  We consider operators of dimension $4+n$ involving two quarks of different flavor. At leading order in the Yukawas, these operators can be made formally invariant under $\gfl$ by introducting of $Y_5$ and $Y_3$:
\be
c^{\alpha\beta}\frac{\mcl{Q}^{\alpha\beta}}{\Lambda^n}  = c\left(Y_5^{\alpha\beta} + c_3Y_3^{\alpha\beta} + c_1\mbb{I}\right)\frac{\mcl{Q}^{\alpha\beta}}{\Lambda^n}  \,,
\ee
where a singlet term needs to be taken into account. Unfortunately, for generic $c_{5,3,1} \sim 1$, operators of this type, such as the magnetic moment operator
\be
\mcl{Q}^{\alpha\beta} = \bar{Q}_L^\alpha \sigma^{\mu\nu} D^\beta_R HF_{\mu\nu} \,,
\ee
lead inevitably to relatively large contributions to flavor changing neutral currents, thus requiring $\Lambda \gtrsim 10^3$ TeV. This is because the combination $c^{\alpha\beta}$ is in general not diagonalized by the transformation to the physical basis in Eq.~\eqref{basis-change}. 

A sufficient condition for $c^{\alpha\beta}$ to be diagonalized by  Eq.~\eqref{basis-change} is
\be
c_3 = \beta_3\,,\quad c_1 = \beta_d \,. \label{coef-rel}
\ee
These relations would occur at least approximately if Flavorspin $SU(2)_V$ were a remnant symmetry of a larger flavor symmetry group, as discussed above. Hence, the same mechanism that guarantees $\mcl{U}_{CKM} \sim \mbb{I}$  could provide the  means to bring the scale of flavor violation closer to the EWSB scale. 


\section{Conclusion}


If we allow ourselves to speculate, the fact that the $CP$-conserving SM solution appears to be close to the boundary in Fig.~\ref{th13th23-zoom} suggests that there is a symmetry specifically realized for the observed values of the SM observables \cite{symborder, symborder1, symborder2}. It would be interesting to determine whether such a symmetry exists and what the consequences are if it does. Other challenges for Flavorspin include building models that implement the $SU(2)_V$ flavor symmetry and extending our treatment to the lepton sector.

To conclude, let us note that the most common way of tackling the Flavor Puzzle has been attempting to explain why the low-energy observables have the values they have. This approach is inherently limited by the fact that continuous regions of high-energy parameters map to the same low-energy observables. On the other hand, since $CP$-conserving Flavorspin determines the fundamental Yukawas uniquely, up to discrete degeneracies, it suggests a different way of looking at the Flavor Puzzle. It is not the values of the observables in $\vec{\,o}_{SM}$ that should be explained but rather the values of the parameters in the solutions $\vec{\,p}$. Flavorspin offers a different parametrization of the Flavor Puzzle which may prove useful.

\emph{Acknowledgments.} We are indebted to André de Gouvea and Pilar Coloma for many discussions and comments on the project.  This work is sponsored in part by the DOE grant \#DE-FG02-91ER40684.


\end{document}